\documentclass{aastex}
\usepackage{emulateapj5}

\begin{document}

\submitted{To appear in The Astrophysical Journal Letters}

\title{Optical Spectra of Type Ia Supernovae at \lowercase{$z=$0.46 and $z=$1.2}}
\author{Alison L. Coil\altaffilmark{1}, Thomas Matheson\altaffilmark{1}, 
Alexei V. Filippenko\altaffilmark{1}, Douglas C. Leonard\altaffilmark{1}, 
John Tonry\altaffilmark{2}, Adam G. Riess\altaffilmark{3}, 
Peter Challis\altaffilmark{4}, Alejandro Clocchiatti\altaffilmark{5}, 
Peter M. Garnavich\altaffilmark{6},
Craig J. Hogan\altaffilmark{7}, Saurabh Jha\altaffilmark{4}, 
Robert P. Kirshner\altaffilmark{4}, B. Leibundgut\altaffilmark{8}, 
M. M. Phillips\altaffilmark{9},
Brian P. Schmidt\altaffilmark{10},
Robert A. Schommer\altaffilmark{9}, R. Chris Smith\altaffilmark{9,11}, 
Alicia M. Soderberg\altaffilmark{12}, J. Spyromilio\altaffilmark{8}, 
Christopher Stubbs\altaffilmark{7}, 
Nicholas B. Suntzeff\altaffilmark{9}, and Patrick Woudt\altaffilmark{8}  }

\altaffiltext{1}{Department of Astronomy, University of California, Berkeley,
CA 94720-3411}
\altaffiltext{2}{Institute for Astronomy, University of Hawaii, 2680
Woodlawn Dr., Honolulu, HI 96822}
\altaffiltext{3}{Space Telescope Science Institute, 3700 San Martin 
Drive, Baltimore, MD 21218}
\altaffiltext{4}{Harvard-Smithsonian Center for Astrophysics, 60
Garden St., Cambridge, MA 02138}
\altaffiltext{5}{Departamento de Astronom\'{\i}a y Astrof\'{\i}sica
Pontificia Universidad Cat\'olica, Casilla 104, Santiago 22, Chile}
\altaffiltext{6}{Physics Department, 225 Nieuwland Science Hall, 
University of Notre Dame, Notre Dame, IN 26556} 
\altaffiltext{7}{Department of Astronomy, University of Washington, 
Seattle, WA 98195}
\altaffiltext{8}{European Southern Observatory, Karl-Schwarzschild-Strasse 2,
Garching, Germany}
\altaffiltext{9}{Cerro Tololo Inter-American Observatory,
Casilla 603, La Serena, Chile.}
\altaffiltext{10}{The Research School of Astronomy and Astrophysics,
Private Bag, Weston Creek PO, ACT 2611 Australia}
\altaffiltext{11}{University of Michigan, Department of 
Astronomy, 834 Dennison Bldg., Ann Arbor, MI 48109}
\altaffiltext{12}{Los Alamos National Laboratory, MS D436,
Los Alamos, NM 87545}

\begin{abstract}

We present optical spectra, obtained with the Keck 10-m telescope, of
two high-redshift type Ia supernovae (SNe Ia) discovered by the High-z
Supernova Search Team: SN 1999ff at $z$=0.455 and SN 1999fv at
$z$ $\sim$1.2, the highest-redshift published SN Ia spectrum.  Both SNe
were at maximum light when the spectra were taken. We compare
our high-$z$ spectra with low-$z$ normal and peculiar SNe Ia as well as 
with SNe Ic, Ib, and II.  There are no significant differences between 
SN 1999ff and normal SNe Ia at low redshift.  SN 1999fv appears to be 
a SN Ia and does not resemble the most peculiar nearby SNe Ia.

\end{abstract}

\medskip

\section{INTRODUCTION}

Cosmological tests using high-redshift type Ia supernovae (SNe Ia)
rule out an $\Omega_{M}$=1 universe \citep{gar98,per98} and 
provide evidence for an acceleration of the cosmic expansion 
\citep{rie98,per99}, attributed to a nonzero
cosmological constant ($\Lambda$). The statistical weight of these 
results is high, so the current focus of our High-z Supernova Search 
Team \citep{sch98} is on systematic effects, most of
which can be convincingly tested using SNe at $z\gtrsim$0.8,
regardless of their precise nature.  We expect dimming from
 systematic effects to grow larger with increasing redshift.
However, the effect of a nonzero $\Lambda$ on apparent magnitude can
decrease with increasing redshift, due to the different redshift
dependence of $\Omega_M$ and $\Omega_{\Lambda}$ (see Figure 4 of
Filippenko \& Riess 2000).  Obtaining data for SNe at $z\gtrsim$0.8 is
therefore critical for testing systematic bias versus real
cosmological effects and the presence of a nonzero $\Omega_{\Lambda}$.

In addition to photometry of high-$z$ SNe Ia needed to determine a
luminosity distance for the Hubble diagram, we obtain spectra of our
SNe to measure the redshift, determine the supernova type, and look
for spectral differences between the low-$z$ and high-$z$ SN Ia
populations that may indicate evolution of SNe with time.  Here we
present spectra of two high-$z$ SNe Ia and address uncertainties in
using the spectra to determine redshifts and the SN type.  Details of
the observing program, finder charts, and light curves will be
presented by Tonry et al. (2001).

\section{OBSERVATIONS AND REDUCTIONS} 

SN 1999ff \citep{ton99} was observed with LRIS (Oke et al. 1995) 
on the Keck-II 10-m telescope on 1999 November 8 UT with 0.7$\arcsec$ seeing.
Use of a 1$\arcsec$ slit and the 150 line mm$^{-1}$ grating resulted
in a spectral resolution of $\sim$20 \AA.  The total exposure time was
2200 s, divided into three exposures dithered 10$\arcsec$ along the
slit to reduce the effects of fringing.  The slit was oriented at a
position angle (PA) of 156$^{\circ}$ to include the galaxy nucleus.
SN 1999fv \citep{ton99} was observed with the same setup on 1999
November 10 UT.  The seeing was 0.9$\arcsec$,
and a slit width of 0.7$\arcsec$ resulted in a spectral resolution of
$\sim$16 \AA.  The slit was oriented to PA=252$^{\circ}$ to include a
star.  The total exposure time was 6000 s, taken in four dithered
 frames.  SN 1999fv was also observed on
1999 November 11 UT with the 400 line mm$^{-1}$ grating, with a
1.0$\arcsec$ slit at the same PA and 1$\arcsec$ seeing, in four
dithered exposures of 1000 s each. Standard CCD processing and optimal
spectral extraction were done with IRAF, and we used our own IDL
routines to calibrate the wavelengths and flux of the spectra and to
correct for telluric absorption bands.  The SN 1999fv data from
both nights were combined, after binning to the same resolution.
Only the first two exposures from November 11 were used, as the
last two were overwhelmingly dominated by sky noise.

\section{DISCUSSION} 

Rest-frame spectra of SN 1999ff and SN 1999fv are shown in Figure 1.
The observed wavelength range for SN 1999ff is 4050--7700 \AA,
corresponding to 2785--5290 \AA\ in the rest frame, while for SN
1999fv the observed wavelength range is 6200--9150 \AA,
corresponding to 2830--4180 \AA\ in the rest frame.  SN 1999ff is
located $\sim$2$''$ from the center of a large, bright elliptical
galaxy with $I$=19.1 mag, whereas the SN was $\sim$23 mag.  
To eliminate contamination from the host galaxy, we scaled
and subtracted the host galaxy spectrum from the SN spectrum.  The
high-$z$ spectra have been smoothed with a Savitsky-Golay filter
\citep{press}, a polynomial fit that preserves line features better
than boxcar smoothing.  The spectra of SN 1999ff and SN 1999fv were
heavily smoothed with a 70 \AA\ and 180 \AA\ filter width,
respectively, in the rest frame.

For SN 1999ff we estimate the age based on its spectral
features \citep{rie97} to be --5 $\pm$2 days.  Photometric data of SN
1999ff will be presented by \citet{ton01} and will determine the
true epoch of maximum brightness.  SN 1999fv was significantly fainter
and we were unable to get a reliable spectral age for it, in
part due to the lack of available early-time spectra of low-$z$ SNe Ia
that extend blueward of 4000 \AA.

The redshift of the host galaxy of SN 1999ff is $z$=0.455 $\pm$0.001,
as determined by Balmer absorption lines in its spectrum, and we adopt
this redshift for the SN.  The host galaxy of SN 1999fv was not
visible in our images, and to calculate the redshift of this SN we
cross-correlated the high-$z$ spectrum with several low-$z$ SN Ia spectra
near maximum light (as defined by the $B$-band maximum), 
the ages of which are determined from their light
curves.  Correlating against a sample of 31 spectra with ages
between $-$8 days to +5 days relative to maximum, the redshift of SN
1999fv is $z$=1.17-1.22 for the unsmoothed spectrum, somewhat lower
than the initial estimate of Tonry et al. (1999; $z$=1.23) made 
at the telescope.  Using the same
technique on the spectrum of SN 1999ff, correlating against a sample
of 18 spectra with ages between --7 and --3 days, we find
$z$=0.458$\pm$0.006, consistent with the host galaxy redshift.
We also include galaxy and M-star spectra among our templates for
cross-correlation, which did not fit the high-$z$ spectra well.

The redshift of SN 1999fv has the quoted uncertainty  (1.17--1.22) 
due to the low signal-to-noise (S/N) ratio of the spectrum, and
to its high redshift placing key SN features near bright sky lines at
observed wavelengths $\sim9000$ \AA.  In order to check the validity
of our redshift estimates, we performed blind tests with fake spectra
added to the 2D frames.  We redshifted each of 14 low-$z$ SN Ia
spectra by an arbitrary amount between $z$=0.5 and $z$=1.3, scaled the
flux level down to the signal of SN 1999fv in each 1000 s exposure, 
and added Poisson noise.  We 
added the spectra to the 2D frames (which are dominated by sky emission), 
convolving the signal with a Gaussian across several pixels
perpendicular to the dispersion, imitating seeing effects.  Each 
spectrum was then extracted, and we verified that its S/N ratio as a
 agreed with that of SN 1999fv.  The spectrum
was smoothed heavily with a Savitsky-Golay filter to get an initial
redshift estimate by eye, and subsequently cross-correlated with a
database of low-z spectra to find a quantitative redshift estimate and
uncertainty.  We obtained good results for 12 of the 14 spectra, all
of which matched the input redshift to within the error bars.  For one
spectrum we were unable to estimate the redshift, as the signal
was dominated by sky lines and SN features were not clearly
visible. For another spectrum we obtained two equally likely
redshifts, one of which was the correct input redshift.  In neither
of these two cases were we led to believe an incorrect redshift.
The redshift range for SN 1999fv of z=1.17--1.22 is roughly a 
95$\%$-confidence
level, based on the tests done.  It is important to note that 
the difference in luminosity distance
at z=1.17 and z=1.22 is 0.11 mag (5$\%$) which is much smaller
than our distance uncertainty.  This redshift uncertainty
therefore has negligible impact on our ability to constrain cosmology.

The largest concern with using SNe Ia in cosmological tests is a
possible photometric difference between the low-$z$ and high-$z$
samples.  Spectral comparisons of the distant and nearby SNe could show
subtle evolutionary effects, if present.  If the spectra do not show
differences, this does not prove that the peak luminosities of the SNe
are identical, but it does build confidence that the two samples are
similar.  Spectra of normal, nearby SNe Ia at the same phase are quite
homogeneous \citep{bra93,fil97,rie97}.  As a qualitative comparison
with our high-$z$ spectra, we present in Figure 1 spectra of four
low-$z$ SNe Ia near the same epoch: SN 1989B at both --7 and --1 days
relative to maximum \citep{wel94}, SN 1992A at --1 day \citep{kir93},
and SN 1981B at maximum \citep{bra83}.  The SN 1989B spectra have been
dereddened by $E(B$ -- $V$)=0.32 mag \citep{wel94}.  The --1 day SN
1989B spectrum is a composite of a CTIO spectrum redward of 3300 \AA\
and an {\it International Ultraviolet Explorer} ({\it IUE}) spectrum
from --4 days blueward of 3300 \AA.  The SN 1992A spectrum is a
composite of a CTIO spectrum redward of 3600 \AA\ and a {\it Hubble
Space Telescope} ($HST$) spectrum blueward of 3600 \AA, taken 5 days after 
maximum \citep{kir93}.
For comparison we also show SNe II, Ib, and Ic spectra: SN IIb 1993J
\citep{fil93}, SN Ib 1998dt at maximum \citep{jha98}, and SN Ic 1994I at --4
days, dereddened by $E(B$ -- $V$)=0.45 mag (Filippenko et al. 1995;
see Filippenko 1997 for a discussion of SN subclasses).  The SN 1998dt
spectrum has been smoothed with a 30 \AA\ Savitsky-Golay filter.  We also
show an {\it IUE} spectrum of SN Ib 1983N at --4 days from 2800 \AA\
to 3300 \AA\ \citep{cap95} at the blue end of the SN Ib 1998dt spectrum 
to compare with the high-$z$ spectra.  The SN IIb 1993J spectrum, taken 16
days past discovery and --3 days relative to maximum, resembles 
a relatively normal SN II.

The SN 1999ff spectrum exhibits many of the features seen in the
low-$z$ SN Ia spectra, most prominently the deep \ion{Ca}{2} H\&K
absorption (``a''), the \ion{Si}{2} $\lambda$4130 dip blueshifted to
4000 \AA\ (``b''), \ion{Fe}{2} $\lambda$4555 and/or \ion{Mg}{2}
$\lambda$4481 (``d''), and \ion{Si}{2} $\lambda$5051 (``i'').  There
are several weaker features seen in all of the low-$z$ SN Ia spectra
that are also apparent in the spectrum of SN 1999ff, including a
shoulder blueward of \ion{Fe}{2} $\lambda$4555 (``c''), \ion{Si}{3}
$\lambda$4560 (``e''), \ion{S}{2} $\lambda$4816 (``f''), and
\ion{Fe}{2}/\ion{S}{2} blends (``g'' and ``h'').  All line
identifications (but not the specific letter labels) are from Jeffery
et al.  (1992), Kirshner et al. (1993), and Mazzali et al. (1993).
Unfortunately, the SN 1999ff spectrum does not extend to rest
wavelength 6150 \AA, where the \ion{Si}{2} $\lambda$6355 feature is
prominent in SNe Ia and much less apparent in SNe Ic/Ib.  This line is
the usual diagnostic for distinguishing between SNe Ia and Ib/Ic.  SN
1999ff does not, however, resemble the spectra of SN Ib 1998dt or SN
Ic 1994I, in which the \ion{Si}{2} and \ion{Si}{3} features ``b'' and
``e'' are either very weak or absent.  The SN Ib and Ic spectra also
have a broad peak centered at 5250 \AA\ that is not seen in SN 1999ff
or the SN Ia spectra.  A comparison with the SN IIb 1993J spectrum
shows that while SN 1993J does have \ion{Ca}{2} H\&K (``a'') and
absorption at 3980 \AA\ due to blueshifted H$\delta$ which results in
a double-peak centered just blueward of 4000 \AA, the resemblance with
SN 1999ff redward of 4100 \AA\ is poor.  Very early SN II spectra are
blue and nearly featureless and would be easily classified; they begin
to resemble the SN IIb 1993J spectrum shown here a few weeks after
discovery.

Due to its high redshift, the spectrum of SN 1999fv does not have
broad rest-frame coverage and does not extend redward of
4200 \AA.  SN 1999fv does display the \ion{Ca}{2} H\&K feature (``a'')
as well as \ion{Si}{2} $\lambda$4130 (``b'').  In the UV region
of the spectrum, there are perhaps hints of small rises at 3150 \AA\ and
just blueward of 3000 \AA, as seen in the {\it IUE} spectra
of SN Ia 1992A and SN Ic 1983N. SN 1999fv is bluer than the low-$z$
spectra below 3700 \AA.  Our discovery image for this object does not
reveal an obvious host galaxy, and it
is unlikely that the blue color is due to host galaxy contamination as
SN 1999fv was at least 1 to 1.5 mag brighter than the host.  The SN Ib
1998dt and SN Ic 1994I spectra provide a worse match to SN 1999fv
around 4000 \AA\ than do the normal SNe Ia.  While the spectrum of SN
1999fv has a limited rest-frame wavelength range, the presence of the
\ion{Si}{2} $\lambda$4130 (``b'') feature is very important, as it
results in a broad double-peak in the spectrum centered at $\sim$4000
\AA\ that is not seen in SN Ib/Ic spectra \citep{clo00}, which show a
small dip at this wavelength but lack a strong peak just redward at
$\sim$4100 \AA.  Thus, even for $z\gtrsim$1 SNe with noisy spectra, it
is possible to distinguish features that differentiate SNe Ia from Ib
and Ic events near 4000 \AA.  The SN IIb 1993J spectrum has a
double-peak centered near 4000 \AA, but the widths of the features
blueward of 4200 \AA\ are significantly narrower than in the SN Ia
spectra and the SN 1999fv spectrum.

In general, substantial contamination of the high-$z$ SNe Ia sample with SNe
Ic, Ib, and II seems unlikely.  First, typical SNe Ic, Ib, and IIb are much
fainter than SNe Ia; on average, the $B$-band peak brightness of SNe
Ib and Ic is $\sim$1.5 mag lower, while SNe II are $\sim$1.8 mag lower
(Miller \& Branch 1990; Richmond et al. 1998).  However, Clocchiatti
et al. (2000) have found a very luminous SN Ic, and if such objects
are common at high redshift they might contaminate the sample.  
Second, nearby SNe Ic, Ib, and IIb are rarer than SNe Ia, with a
relative rate of 0.25 to 0.5 \citep{cap97,cap99}, though the
relative rates are likely to increase at high redshift.  Additional
photometric data (light-curve shapes) are useful in ruling out 
contamination from SNe Ic,
Ib, IIb, and II.  We do not believe that
there is confusion with very luminous, peculiar SNe Ic, also known as
``hypernovae,'' in the high-$z$ sample, as their spectra are quite
dissimilar from those of our high-$z$ objects \citep{iwa98,iwa00}.

Although the analysis techniques of Hamuy et al. (1996) and Riess et
al. (1998), in which light-curve shapes are used to derive luminosity
corrections, can be applied to SNe Ia having a range of luminosities
and light-curve shapes, it is of interest to determine whether the
proportion of peculiar high-$z$ SNe Ia differs from that at low-$z$.
In Figure 2 we compare our spectra of SN 1999ff and
1999fv to the two archetypical, very peculiar SNe Ia: overluminous
SN 1991T at --5 days \citep{fil92a} and
underluminous SN 1991bg at maximum \citep{fil92b}.  Our two high-$z$
SNe Ia do not appear to be these kinds of peculiar events, as seen by
the lack of correlation of their spectral features with those of SN 1991T 
and SN 1991bg. SN 1991T lacks \ion{Ca}{2} H\&K as well as 
\ion{Si}{2} $\lambda\lambda$4130, 5051, and exhibits strong \ion{Fe}{3}
lines that are not seen in SN 1999ff and SN 1999fv.  SN 1991bg, on
the other hand, shows prominent \ion{Ca}{2} H\&K absorption but lacks
the usual \ion{Si}{2} lines and has several \ion{Ti}{2} features not
seen in the high-$z$ spectra or in normal SNe Ia.

As a further comparison with peculiar SNe Ia, we show in Figure 2 a
spectrum of SN 1999aa taken at --1 day.  This object is similar to SN
1991T but also exhibits \ion{Ca}{2} H\&K absorption.  Several recent
low-$z$ SNe Ia of this nature have now been classified
\citep{li00,li01}.  While our two high-$z$ SNe Ia are not peculiar in
the sense of SN 1991T-like and SN 1991bg-like events, they do share
some features with the spectrum of SN 1999aa.  However, they agree
best with the normal SNe Ia shown in Figure 1.  We conclude that SN
1999ff is a relatively normal SN Ia, and we do not see any compelling
evidence for SN 1999fv being a very peculiar SN Ia, though wider
wavelength coverage and higher S/N ratios would permit a more
definitive comparison.  In addition, more UV SN Ia spectra are needed
to better understand the nature of peculiar and normal SNe Ia at these
wavelengths.

\section{CONCLUSIONS}

We present spectra of two high-$z$ SNe Ia discovered during our late-1999
High-$z$ SN Search campaign \citep{ton99}.  
A 35 min. exposure of a SN at $z$ $\sim$0.5 yields a
high-quality spectrum with detailed features from which a spectral age
can be derived.  A 130-min. exposure of a SN at $z$ $\sim$1.2 results in
a moderate-quality spectrum with prominent, broad features.  The
redshift of SN 1999ff is 0.455, as given by stellar absorption lines
in the host galaxy, and from spectral features we estimate the age to 
be --5 $\pm2$ days.  For SN 1999fv we use the spectrum to
determine a redshift of 1.17--1.22.  We find that the double-peak in
the spectrum centered at $\sim$4000 \AA\ and the strength of the 
\ion{Si}{2} $\lambda$4130 line can be used as a diagnostic for
distinguishing SNe Ia from SNe Ib and Ic when the usual \ion{Si}{2}
$\lambda$6355 feature is not available, as is often the case for
high-$z$ SNe.  However, moderate-quality
spectra are needed to determine the SN type, especially for 
$z\gtrsim$1 SNe.  We
conclude that SN 1999ff is most consistent with normal SNe Ia, and that
SN 1999fv is a SN Ia and is not peculiar in the sense of SN 1991T and
SN 1991bg-like objects.

This research was supported by NSF grants AST-9417213 and AST-9987438
to A.V.F., by an NSF Graduate Research Fellowship to A.L.C., and by
NASA grants GO-7505 and G0-8177 to the High-z SN Search Team from the
Space Telescope Science Institute, which is operated by the
Association of Universities for Research in Astronomy, Inc., under
NASA contract NAS5-26555.  The Keck Observatory is operated by the
California Institute of Technology, The University of California, and
NASA.  We are grateful to the Keck staff for their assistance with the
observations.  We thank Ryan Chornock and Weidong Li for useful
discussions and the referee for valuable comments.

\begin{figure}
\epsscale{1.0}
\plotone{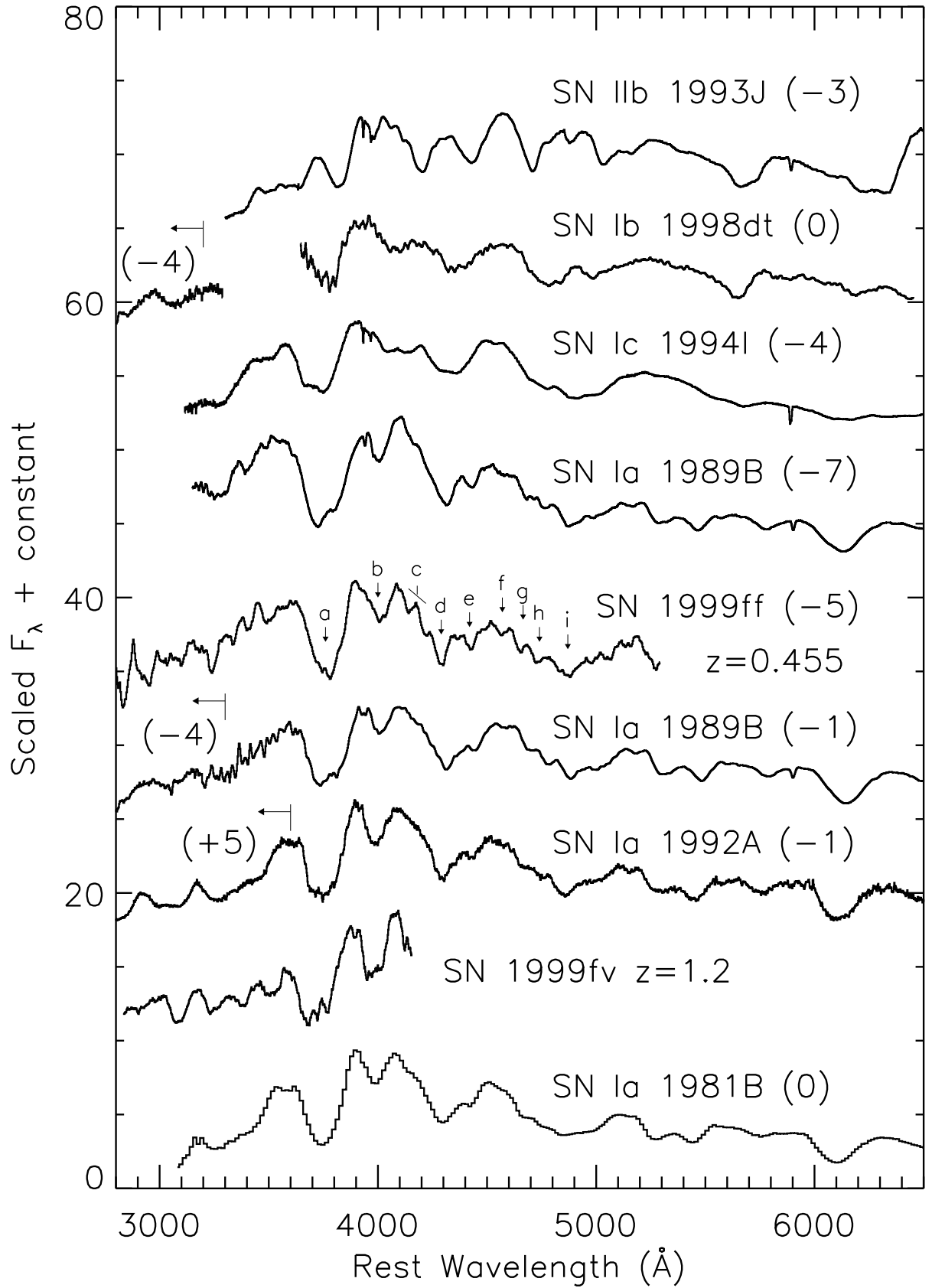}
\caption{Heavily smoothed spectra of two high-$z$ SNe Ia (SN 1999ff
at $z=0.455$ and SN 1999fv at $z=1.2$) are presented along with
several low-$z$ SN Ia spectra (SNe 1989B, 1992A, and 1981B), as
well as SN IIb, Ib, and Ic spectra (SNe 1993J, 1998dt [1983N in the UV], and 
1994I).  The date of the spectra relative to $B$-band SN maximum is
shown in parentheses after each object name.  Specific features seen
in SN 1999ff and labeled with a letter are discussed in the text. The
spectra have been scaled to the same flux level. \label{figure1label} }
\end{figure}

\begin{figure}
\epsscale{1.0}
\plotone{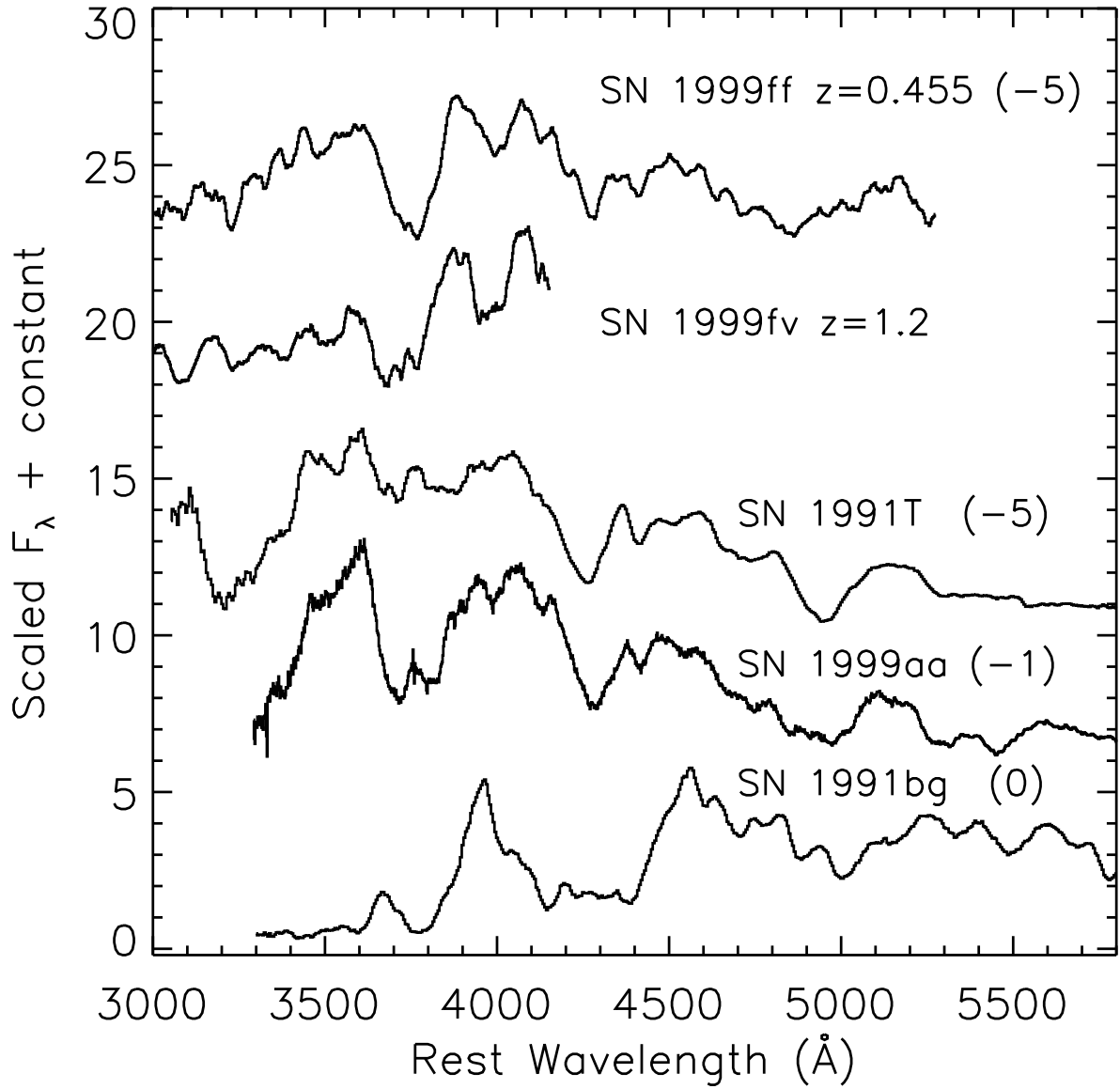}
\caption{Two high-$z$ SN Ia spectra are compared with spectra of
peculiar SNe Ia:  SN 1991T five days prior to maximum light, SN 1999aa one
day prior to maximum, and SN 1991bg at maximum. \label{figure2label} }
\end{figure}

\end{document}